# 3D-Printed Hybrid Liquid-CPCM Cooling Modules for High-Performance Thermal Management of Lithium-Ion Pouch Cells

Xuguang Zhang[a], Michael C. Halbig[b], Amjad Almansour[b], Mrityunjay Singh[c], Meelad Ranaiefar[b], Yi Zheng[a,d]

[a] Department of Mechanical and Industrial Engineering, Northeastern University, Boston, MA 02115, USA

[b] NASA Glenn Research Center, Cleveland, OH 44135, USA

[c] Ohio Aerospace Institute, Cleveland, OH 44142, USA

[d] Department of Chemical Engineering, Northeastern University, Boston, MA 02115, USA

**Abstract**

Efficient thermal management is critical for ensuring the safety, performance, and durability of lithium-ion pouch cells (LIPCs), particularly under high-power operating conditions where conventional battery thermal management systems (BTMS) struggle to balance cooling effectiveness, structural simplicity, and weight. Here, we report a lightweight hybrid BTMS that synergistically integrates active liquid cooling with composite phase-change material (CPCM)-based thermal buffering through a 3D-printed hexagonal architecture. The system is fabricated via a two-step additive manufacturing process that enables sealed CPCM encapsulation and isolated liquid cooling pathways within a single carbon-fiber-reinforced nylon module, effectively eliminating leakage risks while allowing precise geometric control. Hexagonally partitioned CPCM cavities maximize the CPCM–wall interfacial area and shorten internal conduction paths, accelerating latent heat absorption, while embedded serpentine liquid channels provide continuous convective heat removal and prevent CPCM saturation. A nanocarbon-enhanced CPCM is employed to overcome the intrinsic low thermal conductivity of conventional paraffin-based materials. Experimental results using a battery cell simulator demonstrate that the proposed hybrid cooling system reduces peak cell temperature by up to 35 °C, lowering it from approximately 89 °C to 54 °C under representative high heat-flux conditions. Compared with passive CPCM-based cooling, active liquid circulation provides an additional temperature reduction of ~8 °C and significantly stabilizes the thermal response near the CPCM phase-change window. Parametric studies further reveal the coupled effects of coolant flow rate and CPCM cavity partitioning on heat dissipation efficiency and steady-state temperature regulation. Owing to its low added mass (~30% relative to a bare pouch cell), modular design, and manufacturing scalability, the proposed 3D-printed hybrid BTMS offers a practical and robust solution for next-generation high-energy-density LIPC applications, including electric vehicles and stationary energy storage systems.

## 1. Introduction

Lithium-ion pouch cells (LIPCs) are the dominant energy storage technology for electric vehicles (EVs) due to their high energy density and long cycle life [1]. However, battery performance, safety, and longevity are all strongly dependent on temperature regulation [2-3]. Excess heat generation, especially under high current operations such as fast charging, can induce uncontrolled temperature rise, accelerating degradation and even triggering thermal runaway [4-5]. To prevent such outcomes, battery cells must be kept within a narrow optimal temperature range, and temperature gradients across a pack must be minimized to ensure uniform operation and prevent localized overheating [6]. An effective battery thermal management system (BTMS) is therefore essential to dissipate heat and maintain a stable temperature environment for long-term reliability and performance [7-8]. Achieving this goal becomes increasingly challenging as high-capacity battery packs generate a substantial amount of heat, pushing the limits of conventional cooling strategies.

Air cooling is one of the simplest and most widely used BTMS techniques due to its ease of implementation, low cost, and lack of liquid coolant, eliminating leakage risks [9]. Forced-air cooling setups employing fans or blowers can provide some thermal management, but the inherently low thermal conductivity and heat capacity of air limit its effectiveness, particularly in high-power or fast-charging scenarios [10-11]. As a result, air cooling is generally insufficient for next-generation LIPCs with high energy densities, where rapid heat dissipation is crucial to maintaining optimal performance [12-13]. Liquid cooling, on the other hand, provides far superior heat removal capabilities, as liquids have significantly higher thermal conductivity and heat capacity than air [14]. Indirect liquid cooling systems, such as those using cooling plates, tubes, or jackets, have been widely adopted in commercial EV battery packs to tightly control temperature rise and ensure uniformity across cells [15-16]. However, while liquid cooling effectively mitigates excessive heat accumulation, it introduces structural complexity, additional weight, and cost considerations [17-18]. The integration of pumps, reservoirs, and tubing networks increases the system's overall thermal resistance, and potential coolant leakage necessitates robust sealing and insulation, adding to maintenance concerns [19].

Phase-change materials (PCMs) have gained increasing interest as a passive thermal management strategy due to their ability to absorb significant amounts of heat during phase transitions without requiring external power [20-21]. Encasing battery cells in PCMs allows them to buffer temperature spikes and delay excessive heat buildup [22]. While this passive approach enhances thermal stability, PCMs have inherent limitations, particularly their low thermal conductivity, which restricts heat absorption rates and can result in thermal gradients within the material [23]. Once the PCM fully melts, it can no longer absorb additional heat, necessitating active cooling to re-solidify and restore its cooling capacity [24-25]. Additionally, PCM leakage in the liquid phase can compromise system integrity, further limiting standalone PCM-based BTMS applications [26].

Given the limitations of air, liquid, and PCM-based cooling, hybrid BTMS approaches have been proposed as an effective solution to combine the strengths of both active and passive cooling mechanisms [27-28]. By integrating liquid cooling with PCM-based thermal buffering, a hybrid BTMS can leverage latent heat absorption while utilizing the liquid cooling loop to continuously remove excess heat and prevent PCM saturation [29]. Several recent studies have demonstrated the superior performance of such hybrid systems. It is reported that a hybrid BTMS using PCM combined with cooling plates achieved a 24-26% reduction in peak cell temperature compared to natural convection and improved temperature uniformity by up to 56% during charging [30]. Another study showed that a PCM-liquid hybrid system successfully maintained a maximum cell temperature of 31.8 °C during fast charging, whereas a PCM-only system allowed temperatures to rise to 38.4 °C under similar conditions [31]. These findings underscore the advantages of hybrid BTMS designs in maintaining safe battery temperatures during high-power operation [32-33].

Prior hybrid BTMS studies commonly combine PCM matrices with cooling plates, jackets, or immersion schemes to lower peak temperature and improve uniformity, but these assemblies often require separate PCM encapsulation and multi-piece coolant distribution hardware that raise leakage risk, add mass, and limit geometric control [15]. In contrast, the present work leverages an additive two-step process to co-fabricate sealed CPCM cavities and closed liquid pathways within a single CFRN module, enabling leakage-free operation, tunable CPCM-wall interface, and pack-ready modularity. This manufacturing-driven architecture is central to the mechanism we study, because it allows the liquid loop to be tightly coupled to compartmentalized CPCM while preserving precise flow routing and structural integrity.

This work introduces an advanced hybrid BTMS that synergistically integrates liquid cooling with composite phase-change material (CPCM)-based thermal buffering using a 3D-printed hexagonal structure. This novel design provides structural reinforcement while ensuring efficient heat dissipation by maximizing the contact area between the CPCM and cooling channels [34]. The two-step additive 3D printing process effectively isolates the CPCM from the coolant, ensuring no direct contact in between. This manufacturing approach allows for precise geometric control, promoting uniform heat transfer, preventing CPCM leakage, and minimizing material waste [35-36]. The system features embedded serpentine liquid cooling channels adjacent to CPCM-filled hexagonal cavities, enabling continuous heat extraction while preventing CPCM saturation [37]. By leveraging a modular and scalable architecture, this hybrid cooling system offers a lightweight, high-performance BTMS suitable for high-energy-density battery packs in EVs and grid storage applications [38].

**Figure 1** provides an overview of the proposed hybrid cooling system and its structural components. **Figure 1a** depicts the pouch-type LIPC used in this study, with a nominal voltage of 3.7 V, a capacity of 4000 mAh, and an energy density of 174 Wh kg$^{-1}$. The battery dimensions are 114 mm × 50.5 mm × 6 mm, representative of high-capacity cells commonly used in daily applications [39]. **Figure 1b** presents a sectional view of the cooling system, illustrating the spatial arrangement of the CPCM and liquid cooling channels within the hexagonal structure. In the hybrid cooling test system, a polyimide thin film heater (PTFH) was used in place of an LPIC. An aluminum (Al) plate is used to efficiently transfer heat from the LPIC (or PTFH in the test system) to the cooling module. **Figure 1c** displays the fully assembled cooling module without the top panel, highlighting its compact form factor and integrated design. **Figure 1d** provides a top-down schematic of the liquid cooling pathway, demonstrating the structured flow network that ensures efficient convective heat removal. **Figures 1e** and **1f** present two possible battery pack integration configurations for the hybrid cooling system. In the sandwich alignment (**Figure 1e**), cooling modules are positioned on both sides of each battery cell, ensuring efficient thermal management. In contrast, the in-sequence (end-to-end) alignment (**Figure 1f**) optimizes space utilization and enhances coolant distribution efficiency. These configurations demonstrate potential approaches for implementing the proposed hybrid cooling system in battery packs.

The proposed hybrid cooling system effectively mitigates the key challenges associated with conventional BTMS. By combining the thermal buffering capacity of CPCMs with the continuous heat removal capabilities of liquid cooling, the system achieves superior temperature regulation, reducing peak temperatures by up to 35 °C, lowering it from 89 °C to 54 °C compared to standalone cooling methods [40-41]. The hexagonal framework enhances CPCM encapsulation, eliminating leakage risks while ensuring consistent long-term performance [42]. Additionally, the integration of nano-carbon additives into the CPCM significantly improves thermal conductivity, accelerating heat absorption and dissipation rates [43]. Compared to existing BTMS solutions, this novel hybrid approach offers an efficient, lightweight, and scalable design that meets the increasing thermal management demands of high-energy-density battery systems [44-45]. By addressing the limitations of traditional cooling technologies and leveraging advanced additive manufacturing techniques, this hybrid BTMS represents a significant step forward in LIPCs' thermal management, paving the way for safer and more efficient energy storage solutions [46]. Future work will focus on optimizing CPCM composition, exploring alternative coolant fluids with enhanced thermal properties, and refining the hexagonal lattice design to further enhance cooling efficiency and pressure drop minimization [47-48]. The development of intelligent BTMS incorporating real-time temperature monitoring and adaptive cooling strategies will be critical for next-generation high-performance battery technologies [49-50].

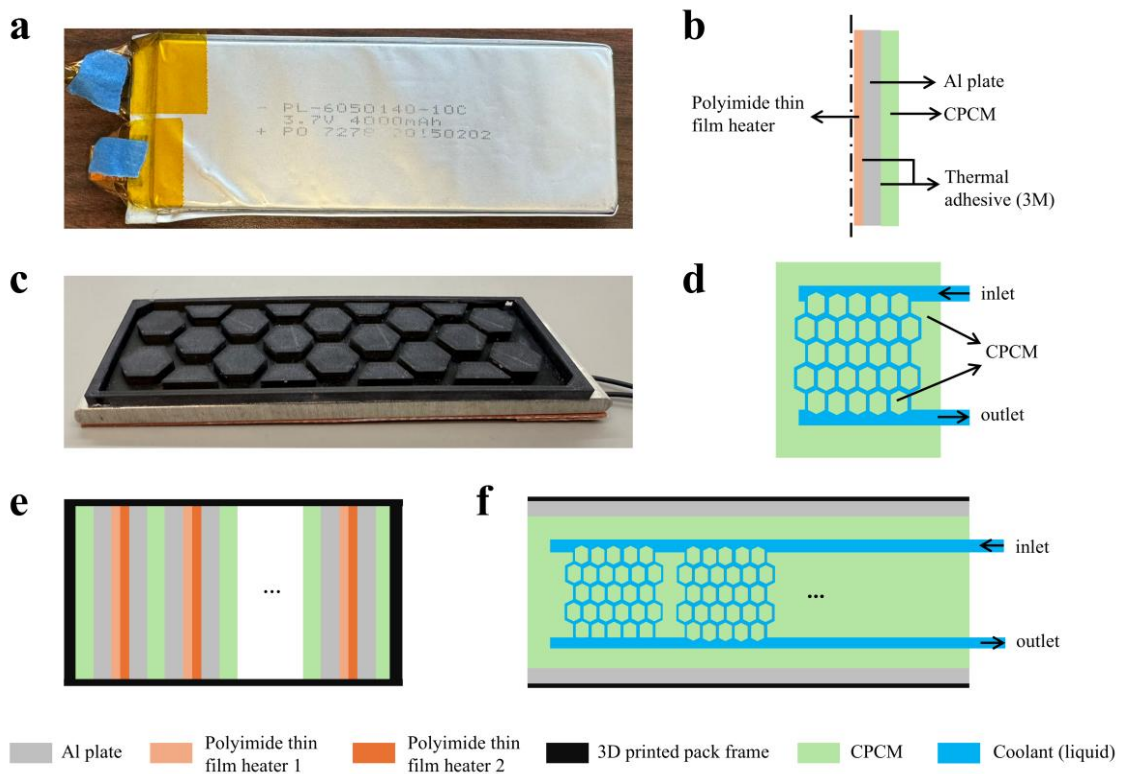

**Figure 1.** Conceptual design of the hybrid cooling system. a) Photograph of the pouch-type lithium-ion battery cell with a nominal voltage of 3.7 V, a capacity of 4000 mAh, and an energy density of 174 Wh kg$^{-1}$. Battery dimensions: 114 mm × 50.5 mm × 6 mm. b) Sectional view of the battery cooling system in which a PTFK is used in place of the LCIP. c) Assembled cooling system without the top panel. d) Top view illustrating the liquid cooling path. e) Schematic representation of the sandwich alignment for multiple battery cells. f) Schematic representation of the top view of the in-sequence (end-to-end) alignment for multiple battery cells.

## 2. Materials and Methods

### 2.1. Hybrid Thermal Management Design

In LIPCs, internal heat generation predominantly arises from two distinct mechanisms: irreversible Joule heating and reversible entropy heat. Joule heating results from electrical energy dissipation due to the battery's internal resistance and is always positive, signifying inevitable energy loss as heat. Conversely, entropy heat is related to thermodynamic changes in electrochemical reactions and can manifest as either heat generation or absorption, depending on the battery's operating temperature and electrochemical characteristics.

The total heat generation rate ($\dot{Q}_{total}$) within the battery can be explicitly defined as the sum of these two contributions:

$$\dot{Q}_{total} = \dot{Q}_{joule} + \dot{Q}_{entropy} = I^2 R_i - IT_b\left(\frac{\Delta S}{nF}\right) \quad (1)$$

where $I$ is the current flowing through the battery, $R_i$ is the internal resistance (Ω), $T_b$ represents the battery cell temperature (K), $\Delta S$ denotes the entropy change associated with electrochemical reactions (J mol$^{-1}$ K$^{-1}$), $n$ is the number of electrons exchanged in the reaction, and $F$ is the Faraday constant (96485 C mol$^{-1}$).

To effectively manage heat generation, a sophisticated cooling strategy was developed specifically for pouch-type lithium-ion battery cells, employing a hybrid thermal management approach. This system synergistically integrates active liquid cooling pathways with passive thermal buffering through CPCMs to ensure efficient thermal regulation. Specifically, an optimized hexagonal framework fabricated via additive manufacturing (3D printing) enhances geometric efficiency and structural robustness while significantly maximizing the available heat dissipation surface area. This design improves thermal exchange between the CPCM and the ambient environment.

Embedded liquid-cooling channels further facilitate active heat extraction, particularly critical under conditions of high thermal loads. Consequently, by combining latent heat storage offered by CPCMs and convective heat dissipation through circulating liquid coolant, the hybrid thermal management system maintains battery temperature stability, enabling rapid and controlled cooling responses under dynamically varying operational demands.

The effectiveness of this hybrid cooling concept can be quantitatively modeled by considering both internal heat generation and external convective cooling. Thus, the net battery temperature change rate can be mathematically expressed as follows:

$$mC_p \frac{dT_b}{dt} = \dot{Q}_{total} - hA(T_b - T_{amb}) \qquad (2)$$

where, $m$ is the mass of the battery cell (kg), $C_p$ is the specific heat capacity (J kg$^{-1}$ K$^{-1}$), $h$ is the convective heat transfer coefficient (W m$^{-2}$ K$^{-1}$), $A$ is the effective surface area for convective cooling (m²), and $T_{amb}$ is the ambient temperature (K).

This comprehensive thermal model explicitly illustrates the interactions between the battery's electrical, thermodynamic, and cooling characteristics, underscoring the critical role of hybrid cooling systems in optimizing battery performance, ensuring reliability, and prolonging cell lifetime under diverse operational conditions. Radiative heat transfer was neglected because at the relevant temperature window near 60 °C the linearized radiative coefficient is on the order of a few W m$^{-2}$ K$^{-1}$, which is small relative to internal convection and solid conduction in this setup. Thermal contact effects between layers were minimized by thermal paste and mechanical preload and were treated within an effective conductance calibrated against the experimental heat-flux and temperature data. Sensitivity checks with a representative contact resistance changed predicted temperatures by less than a few degrees across the operating range.

## 2.2. Fabrication and Characterization

The hybrid cooling system was fabricated through a fused filament fabrication (FFF) approach using a Markforged Mark Two 3D printer with carbon-fiber-reinforced nylon (CFRN) print filament as the primary structural material. This material was specifically chosen for its superior strength-to-weight ratio, outstanding thermal stability, and chemical resistance, ensuring long-term structural integrity under repeated thermal cycling conditions.

A notable innovation in the fabrication process is the development of a two-step 3D printing methodology, carefully engineered to enable effective integration of CPCMs within defined hexagonal cavities. The selected CPCM employs paraffin wax as the matrix and nano-carbon powders as the secondary particulate material [51-52]. The incorporation of 2 wt.% nano-carbon powder substantially improves the thermal conductivity of the CPCM, effectively addressing the inherent heat-transfer limitations of conventional paraffin-based CPCMs. The CPCM exhibits a phase transition temperature of 55-57 °C, a specific latent heat of 173.4 kJ kg$^{-1}$, and a thermal conductivity of 16.6 W m$^{-1}$ K$^{-1}$. The thermal conductivity reported for the CPCM was measured in the solid state at room temperature. The effective conductivity in the fully liquid state is expected to be lower due to reduced particle contact. Additionally, its density is 861 kg m$^{-3}$ in the solid phase and 778 kg m$^{-3}$ in the liquid phase.

Table 1. Thermophysical properties of paraffin wax and the CPCM.

| Property | Paraffin wax | CPCM |
| --- | --- | --- |
| Density (kg m$^{-3}$) | 861 (solid), 778 (liquid) | 840-870 (estimated) |
| Thermal conductivity (W m$^{-1}$ K$^{-1}$) | ~0.2-0.3 | 16.6 |
| Specific heat (J kg$^{-1}$ K$^{-1}$) | ~2000-2900 | ~2000-2900 |
| Latent heat (kJ kg$^{-1}$) | 170-230 | 173.4 |
| Phase change temperature (°C) | 55-57 | 55-57 |

In the initial fabrication step, the 3D printing process established the foundational structure of CFRN, comprising a bottom substrate integrated with precision-designed hexagonal compartments. These compartments serve a dual function: enhancing mechanical stability while precisely delineating cavities for subsequent composite phase-change material (CPCM) injections. The selection of a hexagonal geometry for these cavities is motivated by both manufacturing efficiency and structural optimization. This design is particularly well-suited for additive manufacturing, as it facilitates uniform layer deposition, minimizes material waste, and reduces the likelihood of defects such as warping or delamination during the printing process. Furthermore, the hexagonal structure represents an evolutionary optimization in nature, exemplified by hexagonal formations, which provide exceptional mechanical strength and space efficiency. The inherent tessellation of hexagons ensures a maximal surface area-to-volume ratio, thereby improving heat transfer performance while maintaining a lightweight and structurally robust framework. This combination of manufacturability, thermal efficiency, and mechanical integrity makes the hexagonal design an optimal choice for CPCM encapsulation in advanced battery thermal management systems. Partitioning the same CPCM volume into smaller hexagonal cells increases the CPCM-wall interface and shortens the internal conduction path, which accelerates the onset of melting while preserving structural integrity. After the structural framework was completed, CPCM was carefully injected into these cavities and allowed to solidify (**Figure 2a**) thereby embedding the CPCM securely within the structural matrix and ensuring robust latent heat storage.

The subsequent 3D fabrication step involved printing the top sealing panel of CFRN along with embedded liquid cooling pathways (**Figure 2h**). In the second step, the top panel is printed while the CPCM is molten (maximum volume) to seal each hexagonal cavity; upon cooling the CPCM contracts, avoiding positive pressure on the seals, and the coolant channels are printed as separate closed passages isolated from the CPCM. This stage ensured complete encapsulation of the CPCM, clearly delineating passive cooling elements (CPCM) from active cooling channels (liquid coolant). Consequently, this two-step fabrication approach effectively optimized the thermal management capabilities of the hybrid cooling system, enabling precise control of heat-transfer pathways, high-quality encapsulation, and structural robustness. This materials-process co-fabrication yields a single, leak-tight CPCM-liquid module without secondary potting or housings and is readily transferable to pack-level assembly with low added mass.

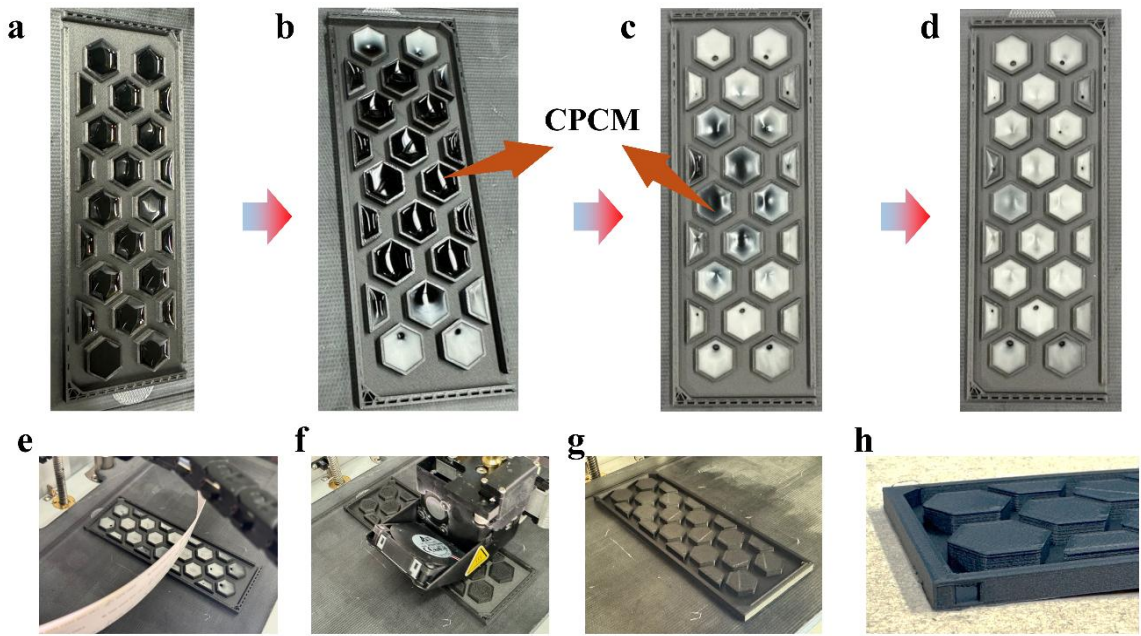

**Figure 2.** CPCM solidification process and top cap fabrication. a)–d) Sequential process of CPCM filling and solidification. e)–h) Continuation of the 3D printing process for the CPCM cap and raised frame.

**Figure 3** presents the experimental validation of the sealing performance of the 3D-printed hybrid cooling system. To rigorously assess potential leakage, two single-unit prototypes were fabricated. **Figure 3a** illustrates the initial condition immediately following water injection, representing the structural integrity test of the CPCM compartments, while **Figure 3b** displays the condition of the prototype after standing for three hours, revealing no observable leakage. This experimental step was crucial for verifying the integrity of the structural cavities designed for CPCM encapsulation. Additionally, given the similarity in viscosity between liquid paraffin wax (dynamic viscosity: $\eta \approx$ 3-5 mPa·s at approximately 60 °C) and water (dynamic viscosity: $\eta \approx$ 0.89 mPa·s at 25 °C), water served as a suitable surrogate fluid for preliminary leakage assessment. **Figure 3** further confirms the structural reliability and functional separation between the CPCM compartments and liquid cooling channels, reinforcing the effectiveness and integrity of the developed two-step 3D printing methodology.

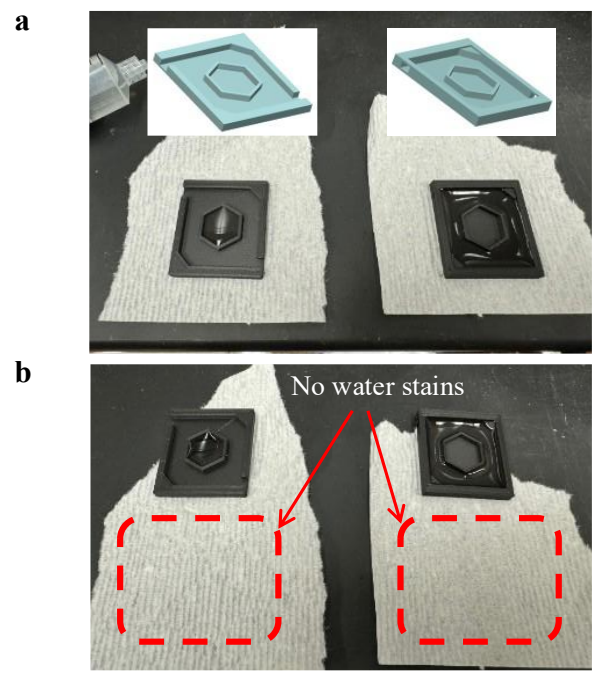

**Figure 3.** Design and water leakage test of the single unit. a) Initial state of the filled water. b) Condition after standing for 3 hours, showing no observable water leakage.

## 2.3. Experimental Setup

Experimental characterization involved various instruments and materials to ensure precise thermal measurement and system performance analysis. A thin-film heater (McMaster-Carr, USA) with dimensions of 152.4 mm× 50.8 mm × 1.78 mm, rated for 120 W power output, provided a controlled heat source to simulate battery thermal loads. Temperature monitoring was conducted using K-type thermocouples (Pico SE000 TC Probe, Pico Technology, UK), connected to a multi-channel thermocouple data logger (Model TC-008, Omega Engineering, USA). Highly conductive aluminum plates (6101 Al, Item #9397T21, McMaster-Carr, USA) served as thermal interfaces and structural supports, chosen for their excellent thermal conduction properties. Active cooling was facilitated by a peristaltic water pump (Masterflex L/S Model 77200-60, Cole-Parmer, USA), supplying a stable coolant flow. Heat flux measurements were obtained using differential-temperature thermopile sensors (FluxTeq PHFS OEM sensors, heat flux range ±150 kW m$^{-2}$, operational range −50 to 120 °C), interfaced via a COMPAQ DAQ data logger from FluxTeq, USA. These sensors provided accurate quantification of thermal transport across critical interfaces, allowing comprehensive evaluation of cooling performance.

Due to safety considerations associated with employing actual LIPCs in thermal studies, a battery cell simulator was designed and fabricated. Specifically, an aluminum plate of 6 mm thickness, integrated with polyimide thin-film heaters on one side and subsequently wrapped with aluminum foil, was manufactured to accurately emulate the power and heat generation characteristics of a single-sided LIPC. The simulator enabled controlled and repeatable generation of thermal loads representative of battery operation. Experimental validation was performed using direct current (DC) voltages ranging from 6 to 10 V and currents from 2.5 to 4.25 A. Precise control and measurement of surface temperature and heat flux were crucial, and sensor locations are illustrated in **Figure 4b** (thermocouples and heat flux sensors). The experimental evaluation identified optimal operating conditions—7 V and 3 A, corresponding to a heating power of 21 W, measured heat flux of 770 W m$^{-2}$, and measured surface temperature of 76 °C—as representative for accurately reproducing the LIPC's single-side thermal behavior during heating and cooling processes.

The experimental configurations utilized to evaluate thermal performance are depicted in **Figure 4a**. The baseline configuration (System 1, **Figure 4b**) consists exclusively of the battery cell simulator, serving as the reference case. System 2 integrates the battery simulator with the hybrid cooling system filled with CPCM, applying a thermal paste between the battery simulator and the cooling system to ensure efficient heat transfer. A representative thermal interface path is on the order of 50-100 μm of thermal paste with k ≈ 3 W m$^{-1}$ K$^{-1}$, giving an areal resistance of about 2-3×10$^{-5}$ m² K W$^{-1}$, which is comparable to or smaller than the bulk CPCM path and is included in the effective conductance used for model-experiment comparison. Building upon System 2, System 3 incorporates a sealed top panel from 3D printing to enhance encapsulation of the cooling structure. Finally, System 4 further refines System 3 by adding an aluminum plate atop the cooling system, with thermal paste applied at the interface, providing reinforced structural integrity and optimized thermal performance. This systematic progression enables comprehensive evaluation and comparison of thermal management effectiveness under progressively enhanced cooling configurations, explicitly emphasizing the performance contribution of the passive cooling component (CPCM) within the hybrid cooling system.

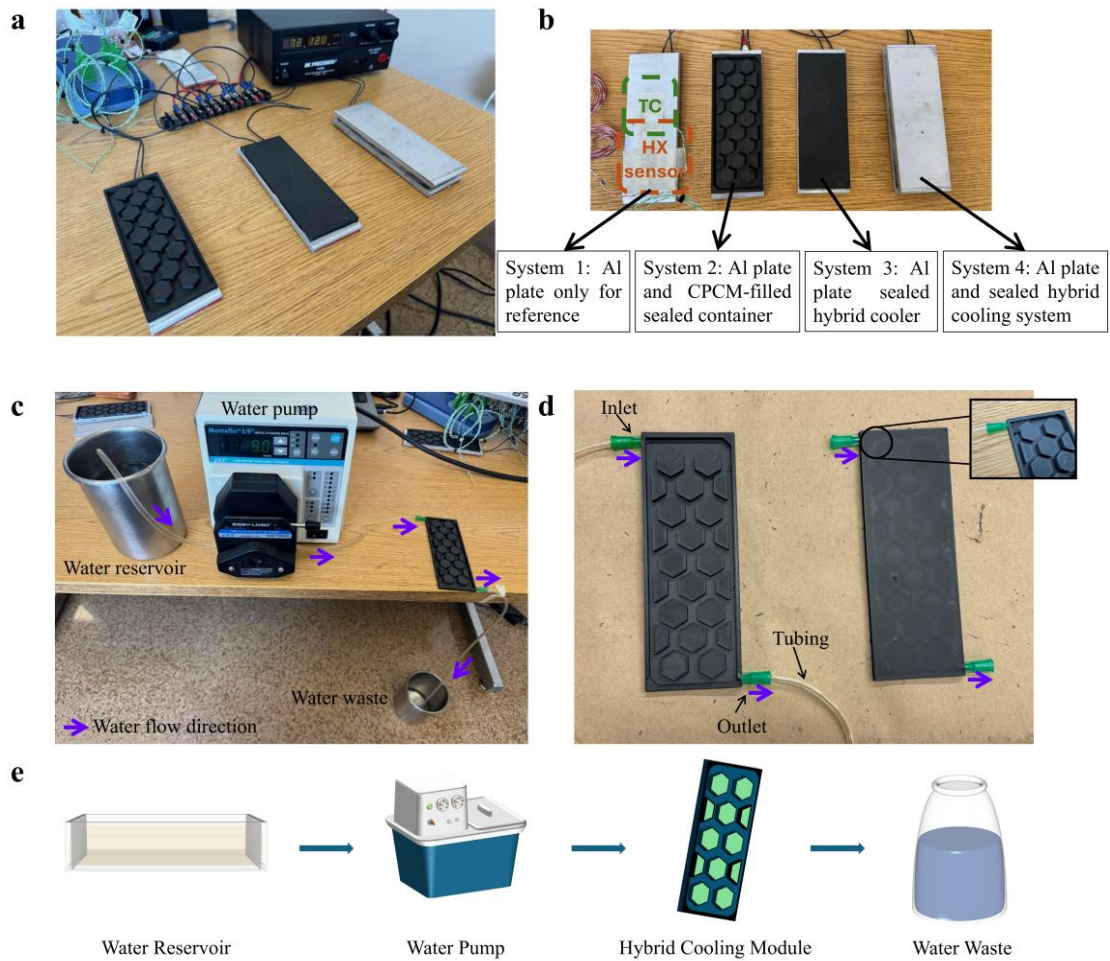

**Figure 4.** Cooling system configuration and experimental setup. a) Overview of experimental components and cooling modules. b) Cooling system configurations for systems 1-4. c) Active cooling setup with water reservoir and pump. d) Water flow path through the hybrid cooling system, with an inset highlighting the water flow behavior. e) Schematic of experimental setup.

The experimental setup for evaluating the active cooling system is shown in **Figure 4c**, and a corresponding schematic is provided in **Figure 4e**. Deionized (DI) water was selected as the liquid coolant due to its adequate thermal conductivity and low viscosity, which facilitates efficient heat transfer while minimizing potential contamination or scaling within the coolant pathways. The cooling process begins with DI water being drawn from a water reservoir into a peristaltic water pump (Masterflex

L/S 77200-60), which regulates and maintains a controlled flow rate. Precise flow rate adjustments were made to ensure uniform coolant distribution and effective heat dissipation across the cooling system.

As depicted in **Figure 4d**, the coolant enters the inlet port of the designed cooling system through the tubing, ensuring a steady and directed flow across the cooling pathways. After passing through the structured flow channels, the coolant exits through the outlet port, where it is collected as waste. The inset in **Figure 4d** indicates that the water flow path remains unchanged after the top seal panel is applied, demonstrating the system's structural integrity and leakage prevention.

A video of the water flow behavior, demonstrating the uniform distribution and controlled movement of the coolant within the system, is provided in the Supporting Material for further visualization. This experimental setup enables precise evaluation of active cooling efficiency, highlighting the role of liquid flow rate in enhancing battery thermal management. The tests were performed at room conditions (ambient 20 °C, relative humidity 50-70 percent) on a closed bench without forced airflow and away from vents or doors. No external airflow was applied; background air speed was negligible.
A summary of the experimental configurations for all tested systems is provided in Table 2.

**Table 2. Summary of experimental configurations for different cooling systems.**

| System | Figure | Structure | Flow Rate (ml min$^{-1}$) | Special Feature |
|---|---|---|---|---|
| S1 | Figure 5a | Only the Al plate | 0 | Reference |
| S2 | Figure 5a | S1 + CPCM + 3D-printed hexagon | 0 | Open structure |
| S3 | Figure 5a | S2 + top panel | 0 | Seal structure |
| S4 | Figure 5a | S3 + 2nd Al plate | 0 | Reinforced Al plate |
| S5 | Figure 5d | S1 + 3D-printed hexagon | 0 | Without CPCM |
| S6 | Figure 6d | S1 + CPCM + 3D-printed small hexagon | 0-24 | Small hexagon |
| S7 | Figure 7a | S4 + sandwich structure | 0-16 | Pack integration simulation |

## 3. Results and Discussion

The thermal behavior of Systems 1-4 (as listed in Table 2) was evaluated through a series of heating and cooling experiments using multiple thermocouples. Eight thermocouples (TCs) were strategically placed at different locations to capture precise temperature data, as demonstrated in **Figure 5a**. These TCs were categorized into three groups based on their positioning: group 1 in direct contact with the battery simulator, group 2 placed within the coolant pathway, and group 3 positioned on the top surface of the hybrid system. Each thermocouple placement was designed to evaluate the battery temperature, heat distribution at the coolant interface, and external surface temperature of the hybrid cooling system. In System 1 (reference system), TC-4 was positioned on the top surface of the battery cell simulator, serving as a baseline for comparison. In System 2, TC-1 was in direct contact with the battery simulator, measuring the temperature of the battery cell, while TC-2 was placed on top of the hexagonal container structure to evaluate surface temperature distribution. Additionally, TC-3 was located at the bottom of the coolant pathway inside the hybrid cooler, monitoring internal heat dissipation. In System 3, the cooling structure was further enclosed with a 3D printed CFRN top panel, and TC-6 was positioned on top of the hybrid cooler, while TC-7 was placed at the bottom of the coolant pathway inside the hybrid cooler, providing additional insights into internal thermal behavior. In System 4, which introduced an aluminum top plate for enhanced heat spreading, TC-8 was positioned on the top surface of the aluminum plate, and TC-5 was placed at the bottom of the coolant pathway inside the hybrid cooler to ensure consistency in coolant-side temperature assessment. The polyimide thin-film heater (PTFH) was used as the heat source, uniformly adhered to one side of the aluminum plate, which was heated using the same configuration described in earlier experiments.

By analyzing the heating data in **Figure 5b** for the first category of thermocouples, TC-1 exhibits a significantly lower temperature than TC-4, indicating that the implementation of the hybrid cooling system effectively enhances thermal management, even when considering only the passive cooling effect. In the second category, a comparison of TC-3, TC-7, and TC-5 reveals that TC-5 records the lowest temperature, suggesting that System 4 maintains a more stable coolant temperature from the inlet to the outlet, thereby minimizing thermal fluctuations within the cooling system. This observation also implies that System 2 provides the most pronounced hybrid cooling effect due to the application of liquid coolant. In the third category, a comparison of TC-2, TC-6, and TC-8 shows that System 4 achieves the lowest surface temperature, demonstrating the effectiveness of the aluminum top plate in dissipating heat. However, the collected cooling data (**Figure 5c**) across all configurations exhibit minimal variation, making it difficult to distinguish significant differences in their cooling performance.

**Figure 5d** presents a comparative analysis of the hybrid cooling system with and without CPCM. In both configurations, the coolant pathway was filled with static water to eliminate the influence of active cooling. The system without CPCM corresponds to TC-12, which is positioned at the back of the hybrid cooler, in direct contact with the battery cell, and represents the reference condition. In contrast, the CPCM-integrated system includes TC-9, located at the backside of the hybrid cooler in contact with the battery cell, TC-10, positioned at the top of the CPCM-filled hexagonal container, and TC-11, placed within the coolant pathway inside the hybrid cooler.

The heating response of these configurations is depicted in **Figure 5e**, where a comparison between TC-12 and TC-9 reveals the thermal regulation effect of CPCM. Although the cooling enhancement provided by CPCM is not highly pronounced, it effectively reduces the peak temperature, demonstrating a moderate but beneficial impact on heat dissipation. The corresponding cooling performance, shown in **Figure 5f**, follows a continuous thermal dissipation trend, reinforcing the role of CPCM in delaying temperature rise and promoting passive heat dissipation within the hybrid cooling system.

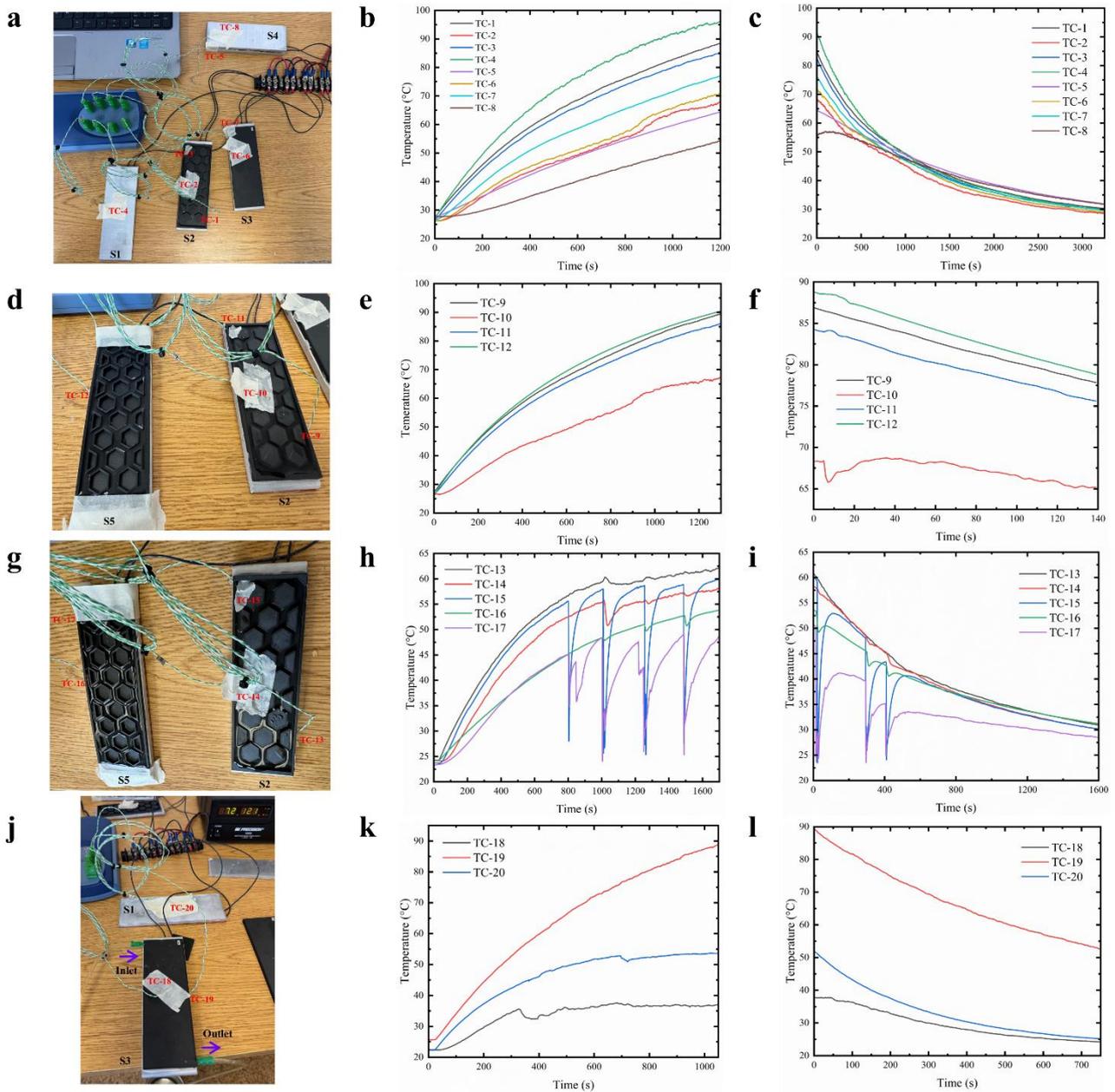

**Figure 5.** Cooling system comparison results. a) Experimental setup with labeled thermocouples. b) Heating data corresponding to (a). c) Cooling data corresponding to (a). d) Comparison of the CPCM effect in the cooling system. e) Heating data corresponding to (d). f) Cooling data corresponding to (d). g) Evaluation of the steady water effect in the cooling system. h) Heating data corresponding to (g). i) Cooling data corresponding to (g). j) Comparison between the reference system and the complete cooling system. k) Heating data corresponding to (j). l) Cooling effect corresponding to (j).

The experiment presented in **Figure 5g** aims to compare the cooling performance of liquid coolant-assisted hybrid cooling with that of pure passive cooling, as previously analyzed in **Figure 5a**. Thermocouple TC-16 is positioned at the backside of the hybrid cooler system without CPCM, while TC-17 is placed within the coolant pathway of the non-CPCM-filled hybrid cooler to monitor fluid temperature variations. For the CPCM-integrated system, TC-13 is located at the back of the hybrid cooler, TC-14 is positioned on top of the hexagonal container, and TC-15 is embedded inside the hybrid cooler, capturing thermal distribution within the CPCM-enhanced structure.

The heating data in **Figure 5h** illustrate a significant reduction in peak temperatures when compared to the results in **Figure 5b**. Specifically, the maximum temperatures recorded by TC-13 and TC-16 demonstrate a substantial temperature drop, with values approximately 33 °C lower than those observed in the pure passive cooling system (TC-4 and TC-1 from **Figure 5b**). In the pure passive cooling configuration, the peak temperature ranges between 90 °C and 95 °C, whereas in the hybrid cooling system, the temperature is notably reduced to a range of 57 °C to 62 °C, underscoring the effectiveness of liquid coolant in mitigating excessive heat buildup. The fluctuations observed in the recorded temperature data are attributed to water evaporation within the cooling system. To maintain a stable water level, additional water was periodically introduced to counteract evaporation losses. The refilling process was halted once the evaporation rate naturally decreased, as reflected in the heat dissipation data in **Figure 5i**. As the temperature gradually declined, evaporation effects diminished, leading to a more stable cooling performance over time.

The experiment presented in **Figure 5j** aims to validate the effectiveness of dynamic water flow in the hybrid cooling system and to compare its cooling performance against static water cooling. The volumetric flow rate of the coolant was precisely controlled at 8 ml min$^{-1}$ to assess its influence on heat dissipation. TC-18 was placed on the top surface of the hybrid cooler, while TC-20 was positioned in direct contact with the battery cell within the hybrid cooling system. For comparison, a reference system without active water flow was also evaluated, with TC-19 placed in contact with the battery cell to monitor its thermal response.

The temperature evolution under steady-state conditions is depicted in **Figure 5k**, where the hybrid cooling system demonstrates a substantial temperature reduction of 35 °C, lowering the battery temperature from 89 °C to 54 °C, a range that ensures safe operational conditions for LIPCs. Furthermore, a comparison between TC-13 (static water cooling) and TC-20 (dynamic water cooling) highlights the significant improvement achieved through active coolant circulation, as the maximum temperature decreases from 62 °C to 54 °C. These results confirm that dynamic water cooling is a highly effective heat dissipation method, offering superior temperature regulation compared to passive or static water-based cooling approaches. Under the calibrated battery-like heat load, the hybrid cooler maintains the cell near the CPCM phase-change window with diminishing returns beyond roughly 8-12 ml min$^{-1}$, while holding near the CPCM window relative to the uncontrolled reference.

To demonstrate the efficiency of the dynamic hybrid cooling system, both the water flow rate and the CPCM cavity layout play a crucial role in optimizing thermal management. The water pump operation range was set between 0 and 24 ml min$^{-1}$, and six specific flow rates were selected for experimental evaluation. To further refine the heat dissipation performance from the standard hexagonal design shown in **Figure 6a,** a modified architecture for the hybrid cooling system was developed. The design retained a hexagonal structure while introducing small hexagonal cavities to explore their impact on thermal behavior, ensuring that the CPCM volume ratio remained constant (**Figure 6d**).

Both hexagonal layouts use the same CPCM volume fraction. The intensive layout partitions the CPCM into smaller cells, which increases the total CPCM-wall interface and shortens the internal conduction path, so each cell reaches melting earlier. At low to medium flow rates, the smaller latent capacity per cell leads to earlier local saturation. Once a cell is fully liquid, the effective thermal conductivity is lower and the local temperature rises until the liquid loop removes the excess heat, yielding a slightly higher steady state than the regular layout. At higher flow rates, enhanced convective removal prevents prolonged saturation, so the larger interface area of the intensive layout reduces the overall thermal resistance, and the temperature approaches the CPCM phase-change range. A modest reduction in effective channel area with the tighter pitch can increase pressure drop and partially offset these gains unless compensated by the pump setting. The heat generation data for the original hybrid cooling system (**Figure 6b**) indicate that at a low water flow rate (2 ml min$^{-1}$), the cell surface temperature continues to rise, approaching 65 °C. However, at higher flow rates (8-24 ml min$^{-1}$), the cell surface temperature stabilizes near the phase-change temperature of 56 °C, suggesting effective heat absorption by the CPCM. In this regime the convective path dominates and additional flow yields progressively smaller temperature gains, defining a practical design window around 8-12 ml min$^{-1}$ for the present geometry and heat load. In contrast, the small hexagonal hybrid cooling system exhibits a slightly higher steady-state temperature than the original design. At a low flow rate (2 ml min$^{-1}$), the temperature stabilizes around 66 °C, while medium flow rates (8-12 ml min$^{-1}$) maintain temperatures near 60 °C. At higher flow rates (16-24 ml min$^{-1}$), the cooling system effectively limits the temperature to within 55 °C, demonstrating an improvement in heat dissipation efficiency. At low to medium flow the smaller CPCM cells melt and locally saturate earlier, which can raise the steady temperature slightly; once convection is strong enough to avoid prolonged liquid-state residence, the larger interface area of the partitioned CPCM lowers the overall thermal resistance. The cooling behavior observed in **Figure 6c** exhibits complexity in predictability, as it is significantly influenced by the water flow rate and the CPCM phase-change dynamics. In general, as shown in **Figure 6f**, the cooling rate increases with higher water flow rates, indicating enhanced heat dissipation. However, the latent heat capacity of CPCM plays a crucial role in this process. A larger CPCM volume possesses greater latent heat storage, requiring a longer duration to complete the phase transition under the same heat transfer conditions. This phase change hysteresis leads to delayed thermal response, making the overall cooling performance more difficult to predict. Consequently, optimizing both CPCM size and water flow rate is essential for achieving efficient and stable thermal management in hybrid cooling systems.

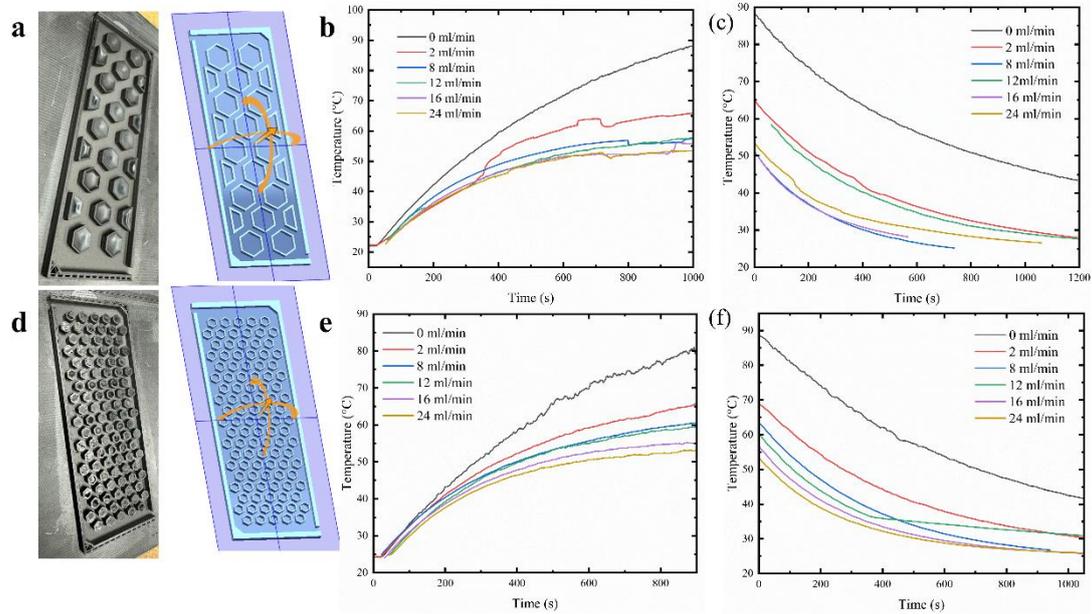

**Figure 6.** Comparison of two different hexagon CPCM-filled designs with variable water flow rates. All thermocouples were positioned in direct contact with the aluminum plate, centered within each cooling system to ensure consistent thermal measurements. a) Regular hexagonal design with a water flow rate ranging from 0 to 24 ml min$^{-1}$. b) Heating data corresponding to (a). c) Cooling data corresponding to (a). d) Intensive hexagonal design with a water flow rate ranging from 0 to 24 ml min$^{-1}$. e) Heating data corresponding to (d). f) Cooling data corresponding to (d).

A new sandwich-structured cooling system has been developed, building upon the design principles of Hybrid Cooling System 4. As shown in **Figure 7a**, this sandwich configuration integrates an additional PTFH on top of the aluminum plate, effectively simulating the thermal behavior of a single hybrid cooling system within a battery pack, similar to the multi-cell arrangement demonstrated in **Figure 1e**. This experimental setup facilitates a comprehensive comparison of temperature distributions across the CPCM cavities and water pathways under varying water flow rates. Specifically, a TC for CPCM was positioned on top of the CPCM cavity, while a water pathway TC was placed on top of the coolant channel. The system without a hybrid cooler serves as the reference configuration for benchmarking performance.

The heating and cooling performance aligns with prior expectations; however, a notable new observation emerges from the heat generation data in **Figure 7b**—the CPCM cavity exhibits slightly higher temperatures compared to the water pathways, a trend consistently maintained across all water flow rates. This uniform temperature distribution across the system confirms the sealed nature of the hybrid system, effectively eliminating any influence from evaporation effects. Despite static water having a significantly lower thermal conductivity than CPCM, the dynamic behavior of flowing water effectively enhances its effective thermal convection, surpassing that of CPCM. This phenomenon underscores the importance of fluid motion in augmenting heat transfer efficiency, providing valuable insights for optimizing hybrid cooling systems in practical battery applications.

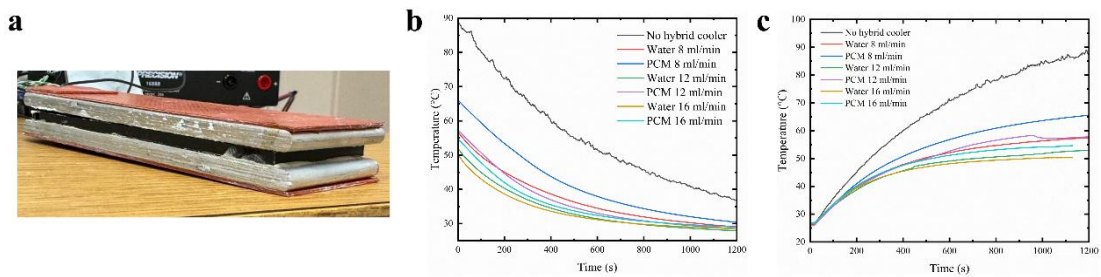

**Figure 7.** Sandwich structure with different water flow rates. a) Display of the sandwich structure. b) Heating data. c) Cooling data.

The mass of the hybrid module was also quantified to evaluate the performance-weight trade-off. Relative to a bare LIPC of 91.67 g, the added mass is 10.43 g for the structure only (~11%), 17.62 g with CPCM and caps (~19%), 23.37 g with CPCM and the top cover (~26%), and 27.48 g when CPCM, top cover, and coolant are included (~30%).

## 4. Conclusion

The study demonstrated that the hybrid cooling system, combining CPCM modules with active liquid cooling, markedly improves heat dissipation and temperature control in LIPCs. Under high thermal loads, the hybrid system reduced peak cell temperatures by about 33 °C compared to a purely passive CPCM-based cooling baseline. In the passive setup, cells reached ~90-95 °C, whereas the CPCM-integrated liquid cooling kept maximum temperatures around 57-62 °C, well below critical thresholds. Furthermore, introducing active liquid circulation (dynamic cooling) led to an additional 8 °C drop in peak temperature relative to static coolant conditions. Overall, the dynamic hybrid cooling held the battery at a safe ~54 °C even when the uncontrolled reference case climbed to ~89 °C, achieving an impressive total reduction of 35 °C in steady-state temperature. These results underscore a significant enhancement in cooling efficiency and thermal regulation, effectively preventing the excessive heat buildup that can occur during fast charging/discharging or high ambient temperatures. The system also maintained more uniform cell temperatures, which is critical for battery longevity and safety, by minimizing hot spots and distributing heat evenly through the hexagon liquid cooling network.

The weight assessment underscores the lightweight nature of the proposed hybrid cooling system. The LIPC weighs 91.67 g, while the hybrid cooler structure alone contributes only 10.43 g. When filled with CPCM and panels, the total weight increases to 17.62 g, and the inclusion of a top panel further raises it to 23.37 g. With water added, the system's total mass remains relatively low at 27.48 g. This lightweight design ensures effective thermal regulation with negligible impact on overall system mass, making it a viable solution for high-performance battery thermal management.

In summary, the proposed hybrid battery thermal management system, leveraging CPCM-liquid cooling synergy and a 3D-printed hexagonal architecture, achieved substantial improvements in heat dissipation and temperature uniformity. By actively pumping the coolant through channels intertwined with CPCM-filled cells, the system effectively absorbed and removed heat, keeping LIPC temperatures well below critical levels even under aggressive thermal conditions. The innovative hexagonal design proved integral to this performance, offering a high surface-area, structurally robust platform for combining latent heat storage with convective cooling. This study's results highlight the viability of hybrid cooling strategies in meeting the thermal challenges of high-power batteries, outperforming traditional single-mode (either passive or active only) cooling approaches.

The work also opens several pathways for refinement. Enhancing the CPCM's thermal conductivity, whether through material chemistry or nanostructures, could further accelerate heat absorption. Likewise, ensuring the mechanical durability of the CPCM encapsulation will be crucial for real-world battery pack deployment over long service lifetimes. On the active cooling side, integrating advanced coolants or flow techniques could push the cooling efficiency even higher. Taken together, these findings and future directions underscore a promising route toward safer, higher-performing LIPCs. By maintaining cells within optimal temperature ranges, the hybrid cooling system not only prevents thermal runaway risks but also contributes to improved battery lifespan and reliability [53-54]. As electric vehicles and energy storage systems continue to demand greater power density, such advanced thermal management solutions will be key enablers for the next generation of battery technology.

**Supporting Information**

Supplementary data associated with this article are available from the corresponding author on reasonable request.

**Acknowledgments**

This project is partially supported by the National Aeronautics and Space Administration Glenn Research Center Faculty Fellowship and the National Science Foundation through grant number CBET-1941743. The authors would like to acknowledge Dr. Jon Goldsby and Dr. Zachary Tuchfeld at the NASA Glenn Research Center for their discussion of the 3D-printed battery capsules and heat dissipation data analysis.

**Conflict of Interest**

The authors don't have any conflict of interest.

**Author Contributions**

Xuguang Zhang: Resources, Writing – original draft, Writing – review & editing. Michael C. Halbig: Resources, Writing – review & editing. Amjad Almansour: Resources, Writing – review & editing. Mrityunjay Singh: Resources, Writing – review & editing. Meelad Ranaiefar: Resources, Writing – review & editing. Yi Zheng: Resources, Writing – review & editing, Project administration, Supervision.